\documentclass[11pt, a4paper]{article}

\usepackage[utf8]{inputenc}
\usepackage[T1]{fontenc}
\usepackage[margin=1in]{geometry} 

\usepackage{amsmath}
\usepackage{amssymb}
\usepackage[version=3]{mhchem} 
\usepackage{gensymb} 

\usepackage{graphicx}
\usepackage{adjustbox}
\usepackage{booktabs} 

\usepackage{authblk}

\usepackage[colorlinks=true, linkcolor=blue, citecolor=blue, urlcolor=blue]{hyperref}

\usepackage[sort&compress,numbers]{natbib}

\title{Simulations of High Temperature Decomposition of Metal-Organic Frameworks to form Amorphous Catalysts}

\author[1]{Connor W. Edwards}
\author[1]{Oliver M. Linder-Patton}
\author[1]{Jack D. Evans\thanks{Corresponding author: j.evans@adelaide.edu.au}}

\affil[1]{School of Physics, Chemistry and Earth Sciences, Adelaide University, South Australia 5000, Australia}

\date{\today} 

\begin{document}

\maketitle

\begin{abstract}
Metal-organic framework (MOF) derived materials formed through high temperature processes show great potential as catalysts. However, understanding of structure-property relationships between the initial MOF and the resulting MOF-derived catalyst is limited because the amorphous nature of the catalyst challenges standard structural characterization methods. Neural network approaches that learn interatomic potentials from density functional theory offer a promising solution. We simulated the pyrolysis of UiO-66, UiO-67 and MIP-206 using both foundational and fine-tuned machine learned interatomic potentials (MLIPs). To mimic experimental conditions, an atmosphere of \ce{CO2} and \ce{H2} was introduced and the structures were doped with 20 wt\% copper to probe the effect of copper on the structural evolution of MOFs. These simulations provide atomistic insights into gas evolution, metal nanoparticle formation, and linker decomposition that were compared to available experimental data. Overall, this work demonstrates the potential of MLIPs to accurately model high temperature MOF dynamics under experimentally relevant conditions and guide the design of new catalytic materials.
\end{abstract}

\vspace{1em}
\vspace{1em}

\newpage
\section{Introduction}
Metal-organic frameworks (MOFs) are a class of compounds composed of metal nodes connected via organic linkers that extend to create crystalline coordination networks.\cite{10.1016/j.ccr.2020.213319, 10.1021/ar5000314} These frameworks typically exhibit high surface area, large pore structures, and high thermal and chemical stability.\cite{10.1126/science.1230444} MOFs are unique because various metallic and organic components can be interchanged to control physical properties such as the porosity and flexibility of the framework.\cite{10.1126/science.1230444, 10.1038/nature01650} As a result, MOFs are being applied to catalysis, gas storage, drug delivery and energy storage.\cite{10.3390/nano12010098, 10.1016/j.mattod.2017.07.006, 10.1002/anie.201915848, 10.1016/j.ensm.2015.11.005} This variability in structure and application led to MOFs receiving the 2025 Nobel Prize in Chemistry.\cite{nobelprize} Post-synthetic modification of these already useful frameworks by pyrolysis or calcination can produce amorphous MOF-derived catalysts that are capable of outperforming conventional catalysts.\cite{10.1021/acs.chemrev.9b00685} For example, in the conversion of \ce{CO2} to methanol reduced MOF-derived catalysts were reported to have higher selectivity for methanol compared to pristine MOFs.\cite{10.1021/acsami.5c10085}

Zirconium-based MOFs, including UiO-66 and MIP-206 are established supports that can stabilize metal nanoparticles to produce high-performing MOF-derived catalysts.\cite{10.1039/C7QM00328E, 10.1021/acsami.5c10085} Compared to conventional metal-oxide supports, zirconium-based MOFs reduce copper agglomeration and minimize deactivation of the catalyst.\cite{10.3390/ma12233902, 10.1016/j.ccr.2023.215409, 10.1039/C7QM00328E} Post-catalyst characterization of these amorphous MOF-derived catalysts has revealed the presence of crystalline domains of \ce{ZrO2} and copper nanoparticles.\cite{10.1021/acsami.5c10085} However, the atomic structures of these amorphous MOF-derived catalysts remain poorly understood because standard structural characterization methods cannot provide precise information.\cite{10.1021/acs.jpcc.2c01091, 10.1039/C7CP08508G, 10.1021/jacs.9b03234} Additionally, the process and mechanisms by which copper-loaded MOFs behave under industrially relevant conditions, such as high pressure and temperature, remains unknown.\cite{10.1021/acsami.5c10085} This impedes insight into structure-property relationships between the crystalline MOF and the catalytically active pyrolysis product. Creating this link will enable the rational design of tailor-made catalysts, providing a foundation for optimizing reactions across diverse environments and operational conditions.\cite{10.1039/D3MA00345K}

Molecular simulations offer one approach to following this structural evolution at the atomic level. Current conventional methods, such as ab~initio molecular dynamics (AIMD), are too computationally expensive to be applied on the temporal and spatial scales required to describe MOF pyrolysis. One approach that has shown success is using machine learned interatomic potentials (MLIPs) to simulate these processes. MLIPs are functions designed to approximate ab~initio potential energy surfaces (PES) and have greater computational efficiency than AIMD.\cite{10.1039/d3dd00236e, 10.1021/acs.jpclett.4c00746, 10.1038/s41563-020-0777-6} This approach has previously been applied to successfully simulate the high temperature dynamics of zinc-based MOFs, identifying phase transitions in zeolitic imidazolate frameworks (ZIFs) and ZIF glass formation.\cite{10.1002/adts.202500514, 10.26434/chemrxiv-2025-71lb4, 10.1039/d3dd00236e} However, generating sufficient accurate training data with AIMD-level accuracy is computationally expensive. Additionally, short timescale AIMD simulations will lead to the emergence of correlated structures within the training dataset, limiting the overall performance of the model at higher temperatures.\cite{10.1002/adts.202500514}

Several pre-trained MLIPs named foundational models, or universal MLIPs, serve as valuable ``off-the-shelf'' tools for atomic simulations.\cite{10.48550/arXiv.2205.06643, Batatia2022mace, 10.48550/ARXIV.2504.06231, 10.5281/zenodo.15587498, 10.1021/acs.jctc.4c00190, 10.1021/jacs.4c14455} The performance of these models are constantly tested by the community and one benchmark, MATBench, currently ranks them from highest to lowest accuracy: fairchem, ORB-v3, SevenNet-MF-ompa, MACE-MPA-0 and MACE-MP-0.\cite{10.48550/arXiv.2308.14920} Foundational models are trained on large open source structural databases, such as MPtrj or OMat24,\cite{10.1038/s42256-023-00716-3, 10.48550/ARXIV.2410.12771} which consist primarily of low-energy trajectories and ground-state structures.\cite{10.1063/1.4812323} This poses a challenge when applying these models to high temperature simulations. MLIPs only retain accuracy within the domain of systems and conditions represented in their training data.\cite{10.48550/arxiv.2405.07105} This issue can be resolved by fine-tuning these foundational models with small amounts of density functional theory (DFT) reference data, representative of a high temperature system.\cite{10.1002/adts.202500514, 10.26434/chemrxiv-2025-71lb4} Rather than generating this data through AIMD simulations, a more efficient method is to run a simulation using a foundational model to create a set of high temperature pyrolysis like products. Reference energies, forces, and stresses can then be calculated for these structures using DFT. This allows the structures to serve as a representative training dataset that captures the full spectrum of thermally activated behavior at substantially reduced computational cost.\cite{10.1002/adts.202500514}

\begin{figure}[htbp]
    \centering
    \includegraphics[]{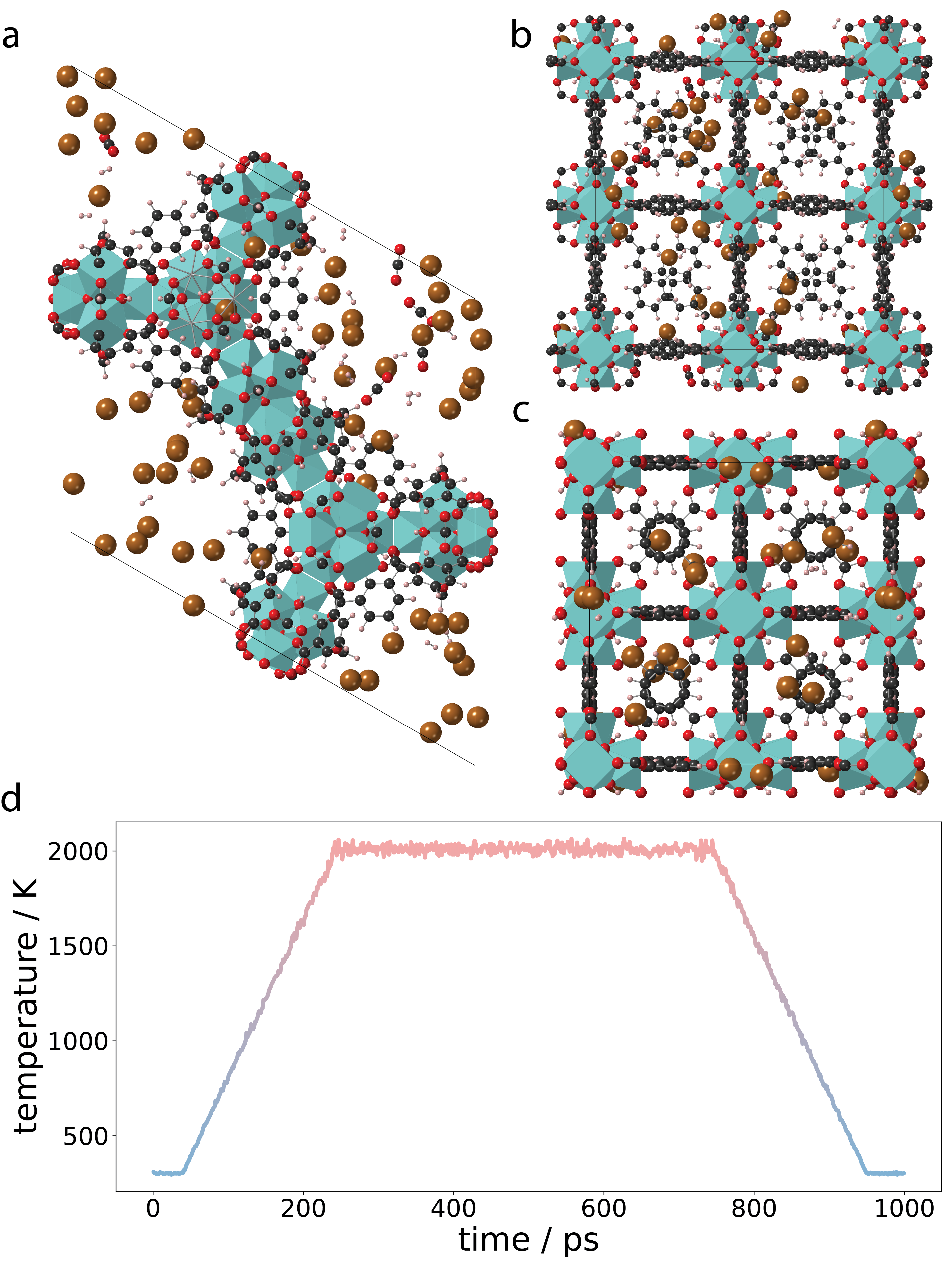}
    \caption{MOF structures of a) MIP-206, b) UiO-67 and c) UiO-66 with a 20~wt\% copper loading and a 40~bar atmosphere of \ce{CO2} and \ce{H2} in a 1:3 ratio. Zirconium atoms are shown in teal, oxygen in red, hydrogen in pink, carbon in black and copper in dark orange. d) An example of the temperature profile for the 1~ns quenches conducted at 2000~K.}
    \label{fig:initial_struct_quench}
\end{figure}

In this work, we employ and fine-tune foundational MLIPs to provide the first atomistic simulations of zirconium-based MOF pyrolysis under catalytically relevant conditions. We simulate zirconium-based MOFs, UiO-66, UiO-67 and MIP-206 (Figure~\ref{fig:initial_struct_quench}) with and without copper doping in a 1:3 \ce{CO2}/\ce{H2} atmosphere at 40~bar. These simulations reveal several previously inaccessible insights, including the atomic-level mechanisms by which copper accelerates linker decomposition and promotes formation of fused carbon ring structures, and the role of linker length in controlling zirconium oxide nanoparticle aggregation. Validation against experimental decomposition products and thermal stability trends demonstrates that fine-tuned foundational models reliably describe high-temperature MOF chemistry, establishing a foundation for computationally guided catalyst design.

\section{Computational Methods}

Reference geometry and cell optimizations were performed for MIP-206, UiO-66 and UiO-67 using the quickstep module of CP2K, version 2023.2.\cite{10.1016/j.cpc.2004.12.014, 10.1063/5.0007045} Valence electrons were described using the triple-$\zeta$ valence polarized basis sets and norm-conserving Goedecker–Teter-Hutter pseudopotentials.\cite{10.1103/PhysRevB.54.1703} The Perdew-Burke-Ernzerhof (PBE) approximation was used for evaluating exchange correlation energy and pair potential interactions were treated by DFT-D3.\cite{10.1103/PhysRevLett.77.3865, 10.1063/1.3382344} These optimized structures were then used to create the MOF models. First, \ce{Cu} atoms were loaded into the framework at 20~wt\%, then \ce{CO2} and \ce{H2} molecules were inserted into the structure (Section~S1 of the Supplementary Information). This was done for a 1:3 \ce{CO2}:\ce{H2} ratio at 40~bar and 300~K. Both \ce{Cu} and gas insertions were conducted using YAFF (version 1.6.0).\cite{yaff} We iteratively attempted to insert a molecule at a random position, rejecting configurations with van der Waals overlap, until the desired number of molecules had been inserted. A series of 1~ns quench simulations were conducted at 2000~K using {MACE-MP-0}. Six simulations were performed across three MOFs: UiO-66, UiO-67 and MIP-206, at copper loadings of 0 and 20~wt\% and a 40~bar atmosphere of \ce{CO2}/\ce{H2}. Each 1~ns quench comprised six steps. First, the structures, loaded with copper and gas, were optimized with MACE-MP-0. The MD simulation then comprised a 50~ps equilibration at 300~K, 200~ps ramp to 2000~K at a rate of 8.5~K\,ps$^{-1}$, 500~ps at 2000~K, 200~ps quench at a rate of 8.5~K\,ps$^{-1}$ and then a further 50~ps equilibration at 300~K (Figure~\ref{fig:initial_struct_quench}). The MD simulations employed a timestep of 0.5~fs in the NVT ensemble using the Langevin thermostat with a friction parameter of 0.01~fs. Each simulation generated 40,000 frames creating a total database containing 240,000 pyrolysis-like products. All calculations in this study were performed using MACE (version 0.3.13),\cite{10.48550/arXiv.2401.00096} SevenNet (version 0.11.2),\cite{10.1021/acs.jctc.4c00190} orb-models (version 0.5.5),\cite{10.48550/arXiv.2410.22570} and fairchem (version 2.4.0).\cite{10.5281/zenodo.15587498} All calculations and simulation trajectories were managed using the Atomic Simulation Environment (ASE) version 3.25.0.\cite{ase-paper}

These initial quench simulations provided the database for training new MLIPs. {MACE-MP-0} and {MACE-MPA-0} were fine-tuned and new MACE-like models were trained ``from~scratch'' using the same features as MACE-MP-0. Three training datasets were constructed by random sampling 108, 180 and 252 structures from each 1~ns quench simulation. This led to three total data sets of 648, 1080 and 1512 structures.\footnote{These datasets were constructed as subsets of a larger preliminary dataset.} Additionally, a test set comprising 10~\% of the training dataset and 1800 validation structures were sampled randomly from the remaining held-out data. These test and validation sets were evenly split across all MOF and copper loading systems. Once structures were sampled, single-point DFT calculations were performed to determine energies, forces and stress tensors for each structure. In addition to data sampled from the 1~ns quenches, additional models, labeled as PT, were trained on 5,000 structures randomly sampled from the original foundational training dataset (MPtrj). Test and validation datasets for these models were the same datasets as those previously specified. In total, six new models were trained.

The fine-tuned {MACE-MPA-0} PT model was used to perform more accurate 1~ns quench simulations for the three MOFs, UiO-66, UiO-67 and MIP-206, at copper loadings of 0 and 20~wt\% under the \ce{CO2}/\ce{H2} atmosphere. Additionally, further simulations were conducted for a $2\times2\times2$ UiO-66 supercell at 0 and 20~wt\% copper loadings under 40~bar atmosphere of \ce{CO2}/\ce{H2}. These quench simulations followed the same temperature profile and had the same specifications as the initial 1~ns simulations. All computational data and analysis scripts supporting this study have been deposited at Figshare (\href{https://doi.org/10.25909/31129939}{10.25909/31129939}). The data are currently under embargo and will be made publicly available upon acceptance.

\section{Results and Discussion}

\subsection{Validating foundational models}
Initially simulations were run using the off-the-shelf MLP MACE-MP-0. These simulations revealed that high temperature decomposition led to the formation of metal nanoparticles, gases such as toluene, xylene and methanol, and an amorphous carbon network that extends through the structure (Section~S2 of the Supplementary Information). However, certain features of these simulations indicate that MACE-MP-0 produces unphysical results. For example, it is experimentally known that copper doped MOFs form copper nanoparticles during pyrolysis.\cite{10.1039/c7ra00115k, 10.1039/C7QM00328E, 10.1021/acsami.5c10085} These simulations did not reproduce this behavior, instead favoring the formation of free copper or small copper clusters. Our observations suggest that MACE-MP-0 is unsuitable for simulating the high temperature dynamics of MOFs. To further investigate this, 300 frames were randomly sampled from each 1~ns quench trajectory and evaluated with DFT to obtain reference energies, atomic forces, and stress tensors. This created an accurate high-temperature MOF dataset which was used to validate a series of six foundational models. The error of each model computed for energy, force and stress is reported in Table~\ref{table:found_mae}. This table also includes the model efficiency and target accuracies. Target accuracies are based on the performance of other MOF trained models and represents a near DFT-level of accuracy.\cite{10.1038/s41524-023-00969-x, 10.1039/d3dd00236e, 10.1038/s41524-024-01427-y} In Table~\ref{table:found_mae}, models are ranked from highest to lowest accuracy, this ordering roughly aligns with their MATBench rankings. This indicates that benchmark performance on MATBench reliably reflects the relative accuracy of foundational models across applications, even when outside the original training domain. In any case, we show all foundational models exceed target error thresholds, suggesting poor suitability for high-temperature decomposition dynamics.

\begin{table*}[htbp]
    \centering
    \caption{MAE for energy, force, and stress and efficiency for a series of foundational models. Efficiency was computed on 1 × NVIDIA A100 GPU (SXM4, 40 GB).}
    \label{table:found_mae}
    \begin{tabular}{llccc}
    \hline
    \textbf{Model} & \textbf{Energy MAE} & \textbf{Force MAE} & \textbf{Stress MAE} & \textbf{Efficiency} \\
    \textbf{ } & \textbf{meV$\,$atom$^{-1}$} & \textbf{meV$\,$\AA$^{-1}$} & \textbf{MPa}  & \textbf{s$\,$atom$^{-1}$$\,$step$^{-1}$}\\
    \hline
    \hline
    Target accuracy & $< 1$ & $< 20$ & $< 200$ & -- \\
    \hline
    fairchem (uma-s-1p1, OMAT) & 11.96 & 183.75 & 352.36 & 4.21e-4\\
    ORB-v3 (conserv. inf.) & 14.22 & 212.07 & 246.69 & 9.61e-5 \\
    SevenNet-MF-ompa & 16.92 & 251.09 & 468.22 & 1.12e-4\\
    MACE-MPA-0 & 20.77 & 300.78 & 314.26 & 2.05e-4\\
    MACE-MP-0 & 25.24 & 308.11 & 288.69 & 3.85e-4\\
    fairchem (uma-s-1p1, ODAC) & 28.61 & 332.62 & 221.83 & 4.25e-4\\
    \hline
    \end{tabular}
\end{table*}

\subsection{Fine-tuning foundational models}
To reduce the high error of the foundational models, a series of models were fine-tuned from MACE-MP-0 and MACE-MPA-0, and additional models were trained "from~scratch" using the MACE architecture. Furthermore, models were trained with 5,000 additional structures randomly sampled from MPtrj (denoted by PT suffix). This step reinforced the existing knowledge of the model during fine-tuning. The previously described validation set was used to validate these trained models and the error is summarized in Table~\ref{table:trained_mae}.

The results show that these models significantly outperform the foundational models and approach target accuracy for energy and stress. Overall, the error in Table~\ref{table:trained_mae} shows ``from~scratch'' models outperform fine-tuned MACE--MPA--0, which in turn outperform fine-tuned MACE--MP--0 models, though all trained models are at a comparable level of error. Table~\ref{table:trained_mae} also shows that the fine-tuned MACE--MPA--0 models are approximately $1.5\times$ more efficient than the ``from~scratch'' and MACE--MP--0 models. Interestingly, inclusion of PT data slightly worsened the performance of the fine-tuned models but improved the ``from~scratch'' models. The ``from scratch'' models initially lacked information about out-of-domain regions of the potential energy surface (PES); the MPtrj data provided this description, enabling the models to represent new structures more accurately. In contrast, the fine-tuned models are constrained to retain more features of their original training data and therefore adapt less effectively to the new high-temperature data. Force error is much higher than anticipated. To investigate this, the force error of each atom was averaged over all frames of the final equilibration and mapped over the atomic structure. Free molecules such as \ce{CO2} and benzene derived species had low force errors on the order of 30~meV$\,$\AA$^{-1}$ (Section~S4 of the Supplementary Information). However, metal sites had a force error around 170~meV$\,$\AA$^{-1}$ and portions of the amorphous carbon framework had errors approaching 200~meV$\,$\AA$^{-1}$. Framework atoms with high error were either part of, or neighboring, a five carbon ring. This suggests the model has below average accuracy in the description of amorphous carbon ring strain, likely because such strained structures are underrepresented in the training data. The elevated force errors at metal sites and strained carbon rings introduce uncertainty in the precise geometries of these regions. However, because the energy and stress errors remain within acceptable bounds, the overall thermodynamics of the decomposition process and the qualitative structural trends (nanoparticle formation, ring statistics, gas evolution) are expected to be reliable. Previous work demonstrated that fine-tuned and ``from~scratch'' MACE models trained on 2000~K data were unstable at 2000~K despite having validation error approaching the target accuracies.\cite{10.1002/adts.202500514}

Metadynamics simulations were used to assess high-temperature stability of various machine learning models (Section~S3 of the Supplementary Information), revealing that Fairchem ODAC and fine-tuned MACE--MP--0  PT exhibited large losses suggesting instability, while other models remained stable beyond the training dataset root-mean-squared-displacement (RMSD) range. However, MD simulations of copper-containing systems using the fine-tuned MACE--MPA--0 model proved unstable at 2000 K despite metadynamics validation. This demonstrates that RMSD-biased metadynamics may not represent a reliable method for testing model stability and better approaches are needed to assess generative performance. Following the failure of the fine-tuned MACE--MPA--0 model, the fine-tuned MACE--MPA--0 PT model was employed. Reinforcement of the original foundational training data provides a better parametrization of out-of-domain regions of the potential energy surface, leading to more stable simulations. This model also has the next best efficiency. The fine-tuned MACE--MPA--0 PT model was used to run a 1~ns simulation for MIP-206, UiO-66 and UiO-67 at both 0 and 20~wt\% copper loadings and a 40~bar atmosphere of \ce{CO2}/\ce{H2}, all of which completed successfully. To validate the generative accuracy of the model, 50 structures were uniformly sampled in time from each trajectory and evaluated using DFT. This gave an energy MAE of 2.46~meV$\,$atom$^{-1}$, force MAE of 105.38~meV$\,$\AA$^{-1}$ and stress MAE of 29.51~MPa. Comparing these results to  Table~\ref{table:trained_mae} shows that fine-tuned MACE--MPA--0 PT model exhibits a generative error consistent with the validation error. Because error is consistently low throughout the simulation, the generated structures are reliable. These results demonstrate that reinforcing the initial training data stabilizes the fine-tuned model by reducing errors arising from out-of-domain MPtrj data. Revisiting this dataset during fine-tuning improves the parametrization of previously misrepresented regions of the PES, enabling the model to better describe out-of-domain structural configurations and remain stable during high-temperature trajectories.

\begin{table}[htbp]
    \centering
    \caption{MAE for energy, force, and stress and efficiency for fine-tuned MACE--MP--0 and MACE--MPA--0 models and ``from~scratch'' models. Efficiency was computed on 1 × NVIDIA A100 GPU (SXM4, 40 GB).}
    \label{table:trained_mae}
    \begin{tabular}{llccc}
    \hline
    \textbf{Model} & \textbf{Energy MAE} & \textbf{Force MAE} & \textbf{Stress MAE} & \textbf{Efficiency} \\
    \textbf{ } & \textbf{meV$\,$atom$^{-1}$} & \textbf{meV$\,$\AA$^{-1}$} & \textbf{MPa}  & \textbf{s$\,$atom$^{-1}$$\,$step$^{-1}$}\\
    \hline
    \hline
    Target accuracy & $< 1$ & $< 20$ & $< 200$ & -- \\
    \hline
    fine-tuned MACE--MP--0 & 3.80 & 102.78 & 31.26 & 3.38e-4\\
    fine-tuned MACE--MP--0 PT & 4.59 & 131.86 & 41.26 & 3.53e-4\\
    fine-tuned MACE--MPA--0 & 8.18 & 91.95 & 31.51 & 1.79e-4\\
    fine-tuned MACE--MPA--0 PT & 2.06 & 109.43 & 32.59 & 2.38e-4\\
    ``from~scratch'' MACE & 3.36 & 95.46 & 30.81 & 3.64e-4\\
    ``from~scratch'' MACE  PT & 3.13 & 95.13 & 29.40 & 4.00e-4\\
    \hline
    \end{tabular}
\end{table}

\subsection{Analysis of fine-tuned 1~ns quench simulations}
\begin{figure}[htbp]
    \centering
    \includegraphics[]{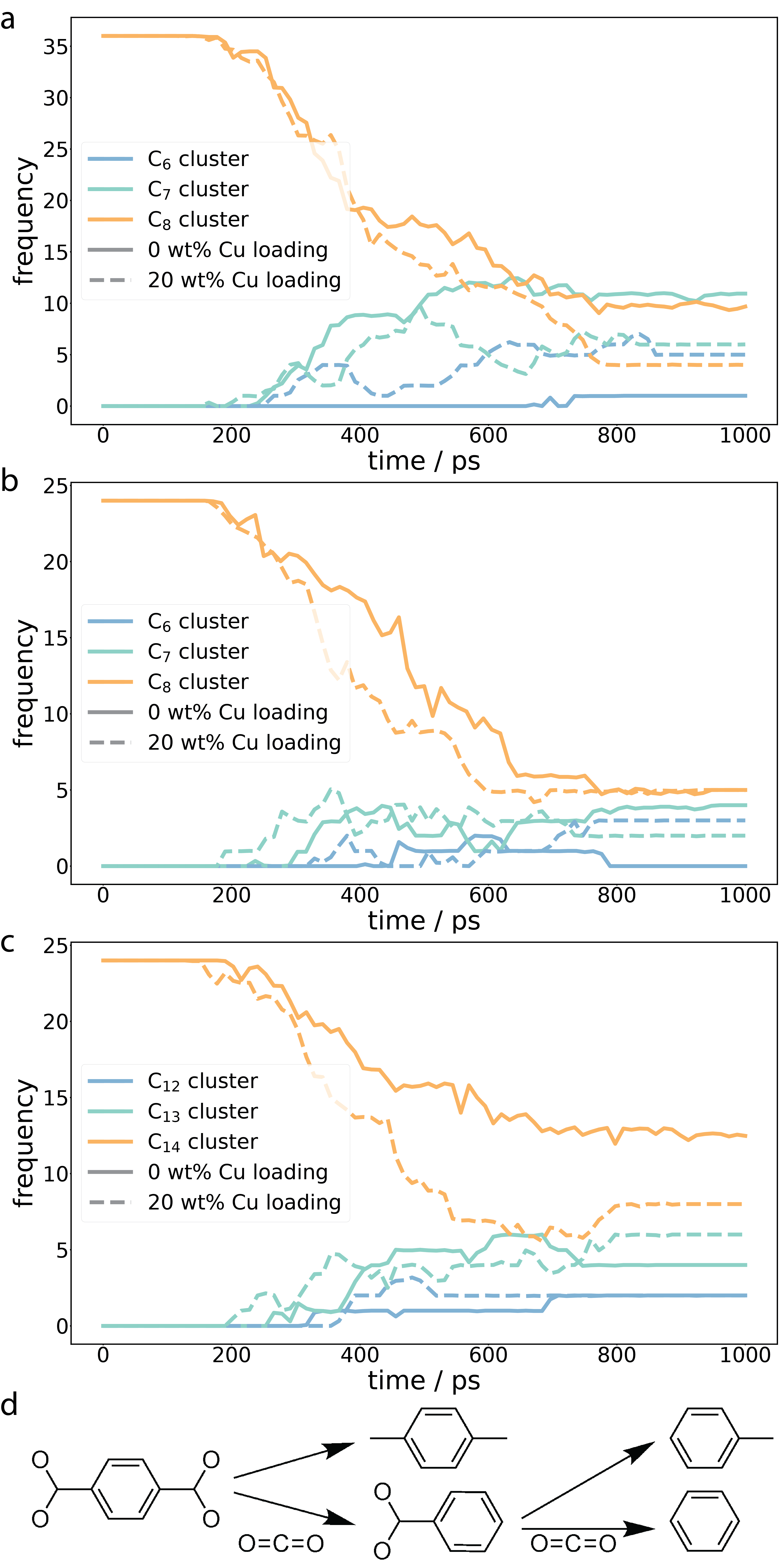}
    \caption{Linker decomposition analysis indicating the sequential loss of one and both carboxylate groups in (a) MIP-206, (b) UiO-66, and (c) UiO-67. (d) Schematic showing decomposition products of the UiO-66 linker.}
    \label{fig:plots}
\end{figure}

The fine-tuned and validated 1~ns simulations were analyzed to follow the formation of zirconium and copper nanoparticles, decomposition of the linker and gas formation. In all simulations, pyrolysis led to linker decomposition as carboxylate groups cleaved to form \ce{CO2}. This process is reflected in Figure~\ref{fig:plots}, where the frequency of carbon groups associated with the initial linker decreases over time (8 carbons for MIP-206 and UiO-66, 14 carbons for UiO-67), giving rise to smaller carbon fragments. This led to the formation of decomposition products such as benzene and \ce{CO2} (Figure~\ref{fig:plots}, Section~S5.2 of the Supplementary Information). Additionally, simulations showed \ce{H2} consumption, resulting from linkers forming reduced organic products, including toluene and xylene (Section~S5.2  of the Supplementary Information). Notably, simulations of MIP-206 also showed methanol formation via the formate pathway, in which the formate binds to metal through the oxygen, reduces to form a methoxy ion and further reduces to methanol. Despite identical \ce{H2} loadings across doped and undoped systems, the doped system initially exhibited less \ce{H2} (Section~S5.2 of the Supplementary Information) resulting from the initial optimization, which dispersed \ce{H2} across the copper clusters. Linker degradation products have been experimentally observed for Cu/Zn-loaded UiO-66 and MIP-206 (10~wt\%). Toluene and para-xylene were detected for UiO-66, and toluene was detected for MIP-206 by sampling the gas mixture via gas chromatography with flame ionization detection (GC-FID) under batch conditions after a 5 hour batch reaction, at 200\degree C, 40 bar, with a 3:1 ratio of \ce{H2} to \ce{CO2}, indicating linker degradation under these catalytic conditions.\cite{10.1021/acsami.5c10085}

Copper accelerated linker degradation in both UiO-66 and UiO-67 and promoted more complete decomposition (Figure~\ref{fig:plots}). Trajectory analysis suggests copper facilitates this process by serving as a hydrogen reservoir that promotes hydrogenolysis of carboxylate C--O bonds, and by catalyzing C--C bond rearrangements that fuse aromatic fragments into extended carbon sheets with mixed C\textsubscript{5}, C\textsubscript{6}, and C\textsubscript{7} rings. In contrast, undoped systems produced isolated aromatic fragments that retained their original six-membered ring character. This copper-promoted decomposition was reflected in increased \ce{CO2} formation (Section~S5.2 of the Supplementary Information). Experimentally, metal doping has been shown to decrease MOF stability during 2~h calcination under atmospheric conditions.\cite{10.1021/acsami.5c10085} This was also observed for UiO-66 and MIP-206 doped with 10~wt\% Cu/Zn, with powder X-ray diffraction (PXRD) analysis demonstrating the loss of MOF long-range order above 350\degree C, well below the reported thermal stabilities of both frameworks.\cite{10.1021/acsami.5c10085, 10.1021/ja8057953, 10.1039/d1ra05411b, 10.1016/j.matt.2020.10.009} 

Not only do the final quenched products retain porosity, but the UiO family also exhibited increased pore sizes (Table~\ref{table:pore_size}). Copper promoted larger pore formation by driving more extensive linker decomposition into gaseous products, increasing the available free volume within the structure. However, as these simulations are limited to a single unit cell, it is unclear whether the pores would interconnect over larger length scales. 

\begin{table}[htbp]
    \centering
    \caption{Comparison of pore size distributions in doped and undoped MOFs at the beginning and conclusion of the 1~ns quench simulations.}
    \label{table:pore_size}
    \begin{tabular}{llccc}
    \hline
    \textbf{MOF \& loading} & \textbf{Initial pore size} & \textbf{Final pore size} \\
    \textbf{ } & \textbf{\AA} & \textbf{\AA} \\
    \hline
    \hline
    MIP-206 0~wt\% & 21.18 & 14.09 \\
    MIP-206 20~wt\% & 21.18 & 14.53 \\
    UiO-66 0~wt\% & 8.62 & 9.13 \\
    UiO-66 20~wt\% & 8.62 & 9.66 \\
    UiO-67 0~wt\% & 12.91 & 13.63 \\
    UiO-67 20~wt\% & 12.91 & 15.19 \\
    \hline
    \end{tabular}
\end{table}

Both the doped and undoped simulations revealed the formation of larger carbon structures (Section~S5.3 of the Supplementary Information). In the undoped systems, these structures were generally connected through single bonds and retained many features of the original linker. This includes the dominance of C6 rings across carbon structures and the retention of carboxylate features (Figure~\ref{fig:final_struc}, Section~S5.1 of the Supplementary Information). In contrast, copper promoted the formation of fused carbon structures composed of interconnected C5, C6, and C7 rings with a more complete loss of carboxylate features (Figure~\ref{fig:final_struc}). These carbon structures tended to be larger than the carbon products formed in the undoped simulations. This behavior was consistent across all MOFs, suggesting that linker and framework topology has a limited impact on amorphous carbon formation in these copper doped systems.

\begin{figure*}[htbp]
    \centering
    \includegraphics[]{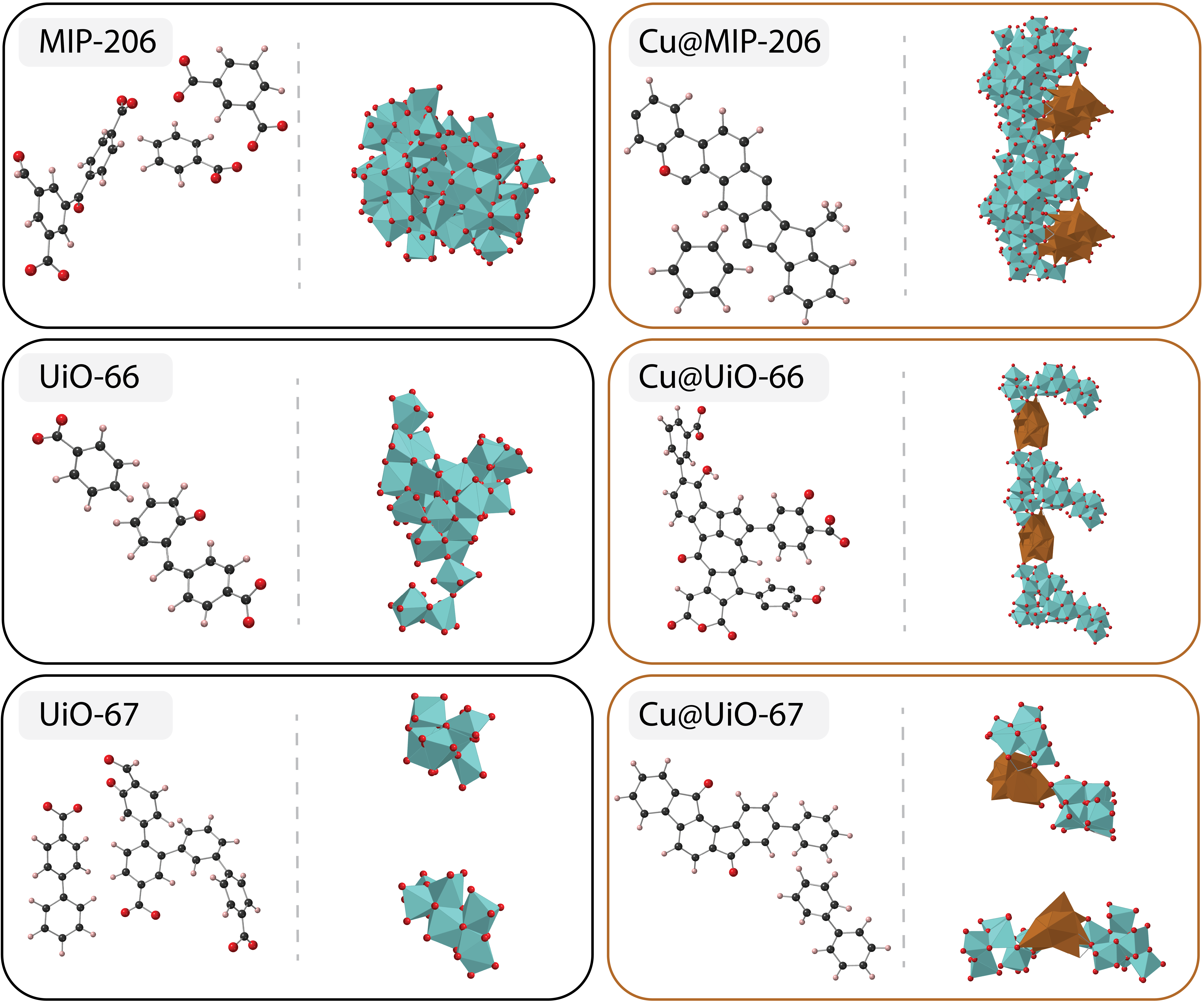}
    \caption{Example structures of amorphous carbon and metal nanoparticles extracted from the final structure of 1~ns quenches of MIP-206, UiO-66, and UiO-67 at 0 and 20~wt\% copper loadings. Zirconium atoms are shown in teal, oxygen in red, hydrogen in pink, carbon in black and copper in dark orange. }
    \label{fig:final_struc}
\end{figure*}

Ramp quench simulations indicated that MIP-206 and UiO-66 formed large zirconium oxide nanoparticles.
In contrast, UiO-67 did not form a larger nanoparticle structure.
Simulations revealed that nanoparticle size remained constant as larger UiO-67 linkers led to greater separation of zirconium nodes and sterically impeded metal mobility, preventing aggregation into larger nanoparticles (Section~S5.4 of the Supplementary Information).
Generally, Zr-MOFs with smaller linkers and zirconium nodes close together, such as UiO-66 and MIP-206, have been experimentally observed to form discrete \ce{ZrO2} nanoparticles after exposure to catalytic conditions.\cite{10.1021/acsami.5c10085} Whereas Zr-MOFs with larger linkers and more widely spaced zirconium nodes (NU-1000) retain long-range order for longer periods, resulting in less \ce{ZrO2} migration and much smaller cluster formation.
In the case of UiO-66 and MIP-206, \ce{ZrO2} domains could be detected via PXRD. Whereas for NU-1000, PXRD analysis did not show the presence of crystalline \ce{ZrO2} domains and electron diffraction was necessary to detect them, indicating that these domains were below 1~nm in size.\cite{10.1021/acsami.5c10085, 10.1021/acsnano.9b05157, 10.1016/j.biotechadv.2013.11.006, 10.3390/min12020205}
This trend mirrors the limited \ce{ZrO2} migration observed by simulations for UiO-67.

Nanoparticle morphology varied across MOFs and with copper doping. In undoped UiO-66 and MIP-206, where the zirconium could migrate freely, it tended to form a single zirconium oxide sphere (Figure~\ref{fig:final_struc}). Although the introduction of copper did not significantly affect the rate of zirconium accumulation, it substantially affected nanoparticle morphology, forming extended pillars of \ce{ZrO2} and copper nanoparticles (Figure~\ref{fig:final_struc}, Section~S5.4 and Section~S5.5 of the Supplementary Information). In UiO-66, these pillars had alternating copper and zirconium oxide domains. In MIP-206, copper was deposited along a zirconium oxide pillar. Even in UiO-67, where no larger nanoparticles formed, smaller copper clusters bridged the zirconium oxide nodes. This suggests that the MOF support promotes copper dispersion throughout the structure.
In previous experiments, the migration of Cu (and Zn) was influenced by the retention of long-range order of the Zr-MOF support, with smaller Cu nanoparticles (detected via PXRD and electron microscopy) observed for the less stable UiO-66 and larger copper agglomerations being seen for the more stable NU-1000.\cite{10.1021/acsami.5c10085} Our simulations could not reproduce exact experimental nanoparticle size trends because copper nanoparticle size was constrained by the copper content in each system, the periodic boundary conditions limited metal mobility despite differences in ligand stability, and the small system size produced artificial pillar-like nanoparticle morphologies. Larger system sizes are necessary to accurately and reliably capture these trends and to avoid finite size artifacts influencing the observed morphologies.

\begin{figure}[htbp]
    \centering
    \includegraphics[]{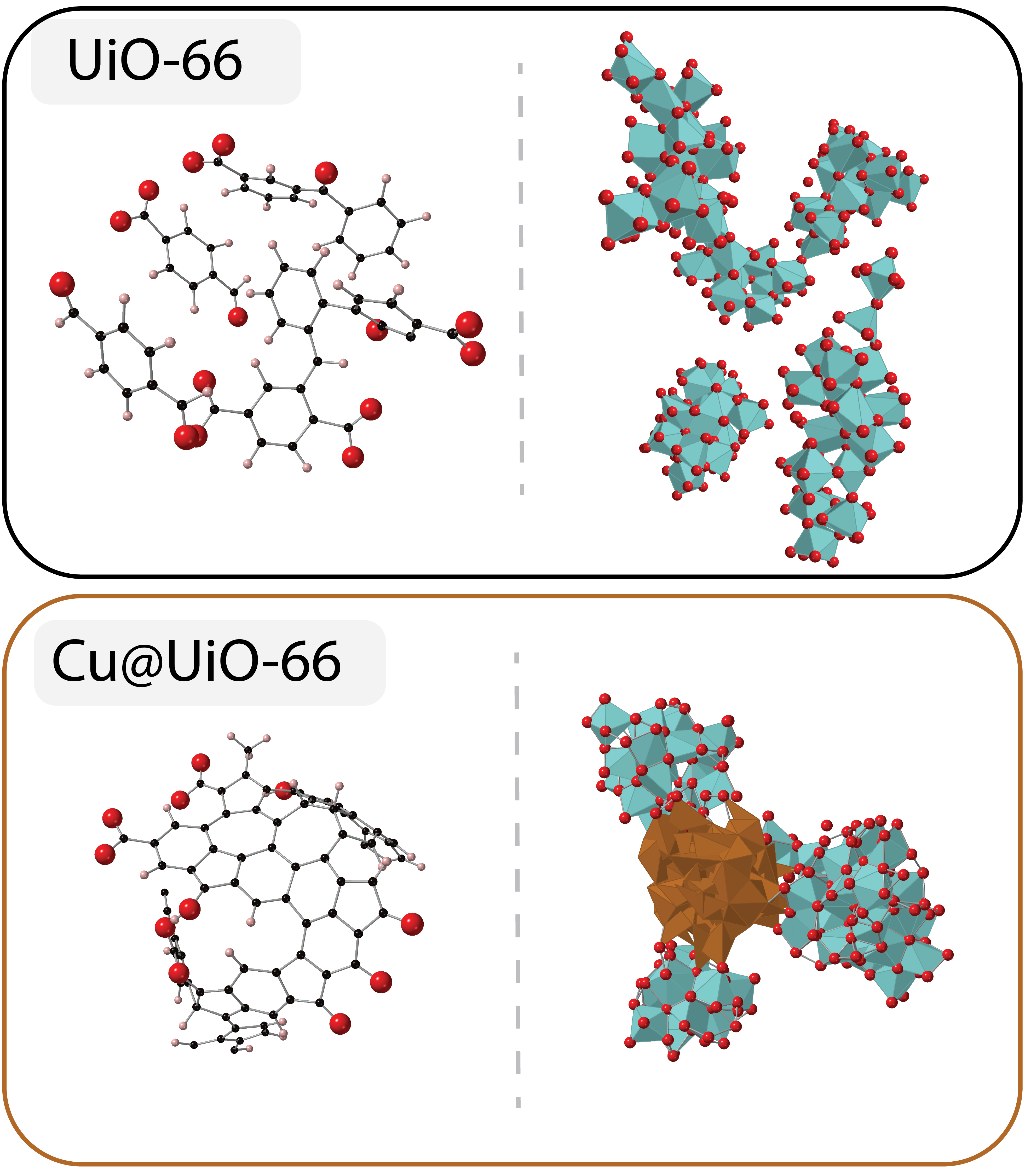}
    \caption{Example structures of amorphous carbon and metal nanoparticles extracted from the final structure of 1~ns quenches of a $2\times2\times2$ UiO-66 simulation at 0 and 20~wt\% copper loadings. Zirconium atoms are shown in teal, oxygen in red, hydrogen in pink, carbon in black and copper in dark orange.}
    \label{fig:final_struc_2x2x2}
\end{figure}

To investigate how these structural characterization changes with system size, 1~ns quench simulations were completed for a $2\times2\times2$ UiO-66 supercell. These simulations followed the same temperature profile as previous quenches and were conducted for both the undoped and 20~wt\% copper doped systems in a 40~bar \ce{CO2}/\ce{H2} environment. Supercell simulations reinforced trends observed in unit cell simulations, including more complete linker decomposition and larger amorphous carbon structures with copper doping, as well as increased gas formation and pore size (Section~S5 and Section~S6 of the Supplementary Information). However, nanoparticle morphology differed significantly: the supercell produced multiple zirconium oxide nanoparticles rather than a single agglomerate, and the pillar-like copper nanoparticle morphology disappeared in favor of zirconium nanoparticles bridged by copper clusters (Figure~\ref{fig:final_struc_2x2x2}). These results demonstrate that nanoparticle morphology predictions are subject to significant size-dependent artifacts. The pillar-like structures observed in unit cell simulations arise from periodic boundary conditions that artificially connect nanoparticles across cell boundaries. Based on the supercell results, we estimate that simulation cells containing at least 10,000--20,000 atoms would be necessary to observe realistic, isolated nanoparticle morphologies and eliminate artificial connectivity. While the chemical trends (decomposition products, gas evolution, ring statistics) are robust across system sizes, morphological predictions from these simulations should be considered qualitative. Future work employing larger supercells or non-periodic boundary conditions\cite{10.1002/adts.201900117} will be necessary to make quantitative predictions of nanoparticle size distributions.

\clearpage

\section{Conclusion and Outlook}
In this work, the formation of amorphous MOF-derived catalysts was investigated through simulations of MOF decomposition under experimentally relevant conditions. Three zirconium-based MOFs, UiO-66, UiO-67, and MIP-206 were simulated under both undoped and copper-doped conditions in a 40~bar atmosphere of \ce{CO2}/\ce{H2}. Foundational models exhibited errors well above target thresholds, producing unphysical processes and unrealistic quenched structures. However, fine-tuning MACE-MPA-0 using high-temperature DFT reference data, combined with 5,000 random structures sampled from MPtrj, produced models capable of generating physically meaningful 1~ns high-temperature trajectories. These simulations enabled characterization of the amorphous MOF-derived catalysts, giving insights into the impact of copper on MOF decomposition.
Copper caused more complete decomposition of linkers and led to larger amorphous carbon sheets forming with C\textsubscript{5}, C\textsubscript{6}, and C\textsubscript{7} ring structures. Additionally, on this atomic scale, the copper led to a larger nanoparticle structure in MIP-206 and UiO-66 where the zirconium oxide served to disperse copper nanoparticles. This dispersion of copper was also seen for UiO-67 where larger zirconium nanoparticles were unable to form. In contrast, undoped systems produced smaller carbon structures that retained features of the initial linkers. This included a dominance of C\textsubscript{6} rings and residual carboxylate groups. The simulations enabled tracking of the evolution of gaseous and formation of other products such as \ce{CO2}, benzene, toluene, and methanol. Supercell simulations confirmed the structural and chemical trends, while revealing nanoparticle morphology is subject to finite-size artifacts that require large-scale simulations.

Our results establish that fine-tuned foundational MLIPs can serve as practical tools for simulating high-temperature MOF chemistry at nanosecond timescales inaccessible to ab initio methods. A key practical finding is that reinforcing fine-tuning data with samples from the original foundational training set improves model stability, while conventional validation metrics (energy/force MAE) and RMSD-biased metadynamics do not reliably predict stability in production simulations. The atomic-level insights obtained here, including the role of copper in accelerating decomposition, the influence of linker length on metal oxide migration, and the identification of specific decomposition products, provide a foundation for understanding structure-property relationships in MOF-derived catalysts. Future work extending these simulations to larger system sizes and additional MOF topologies will be necessary to develop quantitative predictions of nanoparticle morphology and catalytic site distributions.

\end{document}